\documentstyle[aps,epsf]{revtex}

\twocolumn
\newcommand{\be}{\begin{equation}}
\newcommand{\ee}{\end{equation}}

\begin{document}

\title{Structure of Probabilistic Information and Quantum Laws\\
{\small This paper was given as a talk at the conference {\it Foundations of Probability 
and Physics} organized by A. Khrennikov at Vaxjo-University, Vaxjo, Sweden, from Nov.27 to 
Dec.1, 2000. Proceedings will appear sometime in 2001.} }

\author{Johann Summhammer}
\address{Atominstitut der \"{O}sterreichischen Universit\"{a}ten,\\
Stadionallee 2, A--1020 Wien, Austria\\
Email: summhammer@ati.ac.at}
\maketitle
\begin{abstract}The acquisition and representation of basic experimental information 
under the probabilistic paradigm is analysed. The multinomial probability distribution is 
identified as governing all scientific data collection, at least in principle. For this distribution 
there exist unique random variables, whose standard deviation becomes asymptotically 
invariant of physical conditions. Representing all information by means of such random 
variables gives the quantum mechanical probability amplitude and a real alternative. For 
predictions, the linear evolution law (Schr\"odinger or Dirac equation) turns out to be the 
only way to extend the invariance property of the standard deviation to the predicted 
quantities. This indicates that quantum theory originates in the structure of gaining pure, 
probabilistic information, without any mechanical underpinning.
\end{abstract}

\section{Introduction}
The probabilistic paradigm proposed by Born is well accepted for comparing experimental 
results to quantum theoretical predictions \cite{Born}. It states that only the probabilities 
of the outcomes of an observation are determined by the experimental conditions. 
In this paper we wish to place this paradigm first. We shall investigate its consequences 
without assuming quantum theory or any other physical theory. We look at this paradigm as 
defining the method of the investigation of nature. This consists in the collection of 
information in probabilistic experiments performed under well controlled conditions, and in 
the efficient representation of this information. Realising that the empirical information is 
necessarily finite permits to put limits on what can at best be extracted from this information 
and therefore also on what can at best be said about the outcomes of future experiments. At 
first, this has nothing to do with {\em laws of nature}. But it tells us how optimal laws look 
like under probability. Interestingly, the quantum mechanical probability calculus is found as 
almost the best possibility. It meets with difficulties only when it must make predictions from 
a low amount of input information. We find that the quantum mechanical way of prediction 
does nothing but take the initial uncertainty volume of the representation space of the finite 
input information and move this volume about, without compressing or expanding it. However, 
we emphasize, that any mechanistic imagery of particles, waves, fields, even space, must be 
seen as what they are: The human brain's way of portraying sensory impressions, mere 
images in our minds. Taking them as corresponding to anything in nature, while going a long 
way in the design of experiments, can become very counter productive to science's task of 
finding laws. Here, the search for {\em invariant structures} in the empirical information, 
{\em without any models}, seems to be the correct path. Once embarked on this road, the 
old question of {\em how nature really is}, no longer seeks an answer in the muscular 
domain of mass, force, torque, and the like, which classical physics took as such 
unshakeable primary notions (not surprisingly, considering our ape nature, I cannot help 
commenting).  Rather, one asks: Which of the structures principally detectable in probabilistic 
information, are actually realized?

In the following sections we shall analyse the process of scientific investigation of nature 
under the probabilistic paradigm. We shall first look at how we gain information, then how 
we should best capture this information into numbers, and finally, what the ideal laws for 
making predictions should look like. The last step will bring the quantum mechanical time 
evolution, but will also indicate a problem due to finite information.

\section{Gaining experimental information}
Under the probabilistic paradigm basic physical observation is not very different from tossing 
a coin or blindly picking balls from an urn. One sets up specific conditions and checks what 
happens. And then one repeats this many times to gather statistically significant amounts of 
information. The difference to classical probabilistic experiments is that in quantum 
experiments one must carefully monitor the conditions and ensure they are the same for each 
trial. Any noticeable change constitutes a different experimental situation and must be 
avoided.\footnote{Strictly speaking, identical trials are impossible. A deeper analysis of 
why one can neglect {\em remote} conditions, might lead to an understanding of the notion 
of spatial distance, about which relativity says nothing, and which is badly missing in todays 
physics.}

Formally, one has a probabilistic experiment in which a single trial can give $K$ different 
outcomes, one of which happens. The probabilities of these outcomes, $p_1,...,p_K$, 
($\sum p_j = 1$), are determined by the conditions. But they are unknown. In order to find 
their values, and thereby the values of physical quantities functionally related to them, one 
does $N$ trials.
Let us assume the outcomes $j=1,...,K$ happen $L_1,...,L_K$ times, respectively ($\sum L_j 
=N$). The $L_j$ are random variables, subject to the {\em multinomial probability 
distribution}. Listing $L_1,...,L_K$ represents the complete information gained in the $N$ 
trials. The customary way of representing the information is however by other random 
variables, the so called relative frequencies $\nu_j \equiv L_j/N$. Clearly, they also obey 
the multinomial probability distribution.

{\em Examples:} 

\noindent
*  A trial in a spin-1/2 Stern-Gerlach experiment has two possible outcomes. This experiment 
is therefore goverend by the binomial probability distribution. 

\noindent
*  A trial in a GHZ experiment has eight possible outcomes, because  each of the three 
particles can end up in one of two detectors \cite{GHZ}. Here, the relative frequencies 
follow the multinomial distribution of order eight. 

\noindent
*  Measuring an intensity in a detector, which can only fire or not fire, is in fact an experiment 
where one repeatedly checks whether a firing occurs in a sufficiently small time interval. 
Thus one has a binomial experiment. If the rate of firing is small, the binomial distribution can 
be approximated by the Poisson distribution.

We must emphasize that the multinomial probability distribution is of utmost importance to 
physics under the probabilistic paradigm. This can be seen as follows: The conditions of a 
probabilistic experiment must be verified by auxiliary measurements. These are usually coarse 
classical measurements, but should actually also be probabilistic experiments of the most 
exacting standards. The probabilistic experiment of interest must therefore be done by 
ensuring that for each of its trials the probabilities of the outcomes of the {\it auxiliary} 
probabilistic experiments are the same. Consequently, empirical science is characterized by a 
succession of data-takings of multinomial probability distributions of various orders. The laws 
of physics are contained in the relations between the random variables from these different 
experiments. Since the statistical verification of these laws is again ruled by the properties of 
the multinomial probability distribution, we should expect that the {\it inner structure} of 
the multinomial probability distribution will appear in one form or another in the fundamental 
laws of physics.
In fact, we might be led to the bold conjecture that, under the probabilistic paradigm, basic 
physical law is no more than the structures implicit in the multinomial probability distribution. 
There is no escape from this distribution. Whichever way we turn, we stumble across it as the 
unavoidable tool for connecting empirical data to physical ideas.

The multinomial probability distribution of order $K$ is obtained when calculating the 
probability that, in $N$ trials, the outcomes 1,...,K occur $L_1,...,L_K$ times, respectively 
\cite{Feller}:
\be
Prob(L_1, ..., L_K|N, p_1, ..., p_K) = \frac{N!}{L_1!...L_K!}p_1^{L_1}...p_K^{L_K}.
\ee
The expectation values of the relative frequencies are 
\be
\bar{\nu}_j = p_j
\ee
and their standard deviations are
\be
\Delta \nu_j = \sqrt{\frac{p_j(1-p_j)}{N}}.
\ee

\section{Efficient representation of probabilistic information}
The reason why probabilistic information is most often represented by the relative frequencies 
$\nu_j$ seems to be history: Probability theory has originated as a method of estimating 
fractions of countable sets, when inspecting all elements was not possible (good versus bad 
apples in a large plantation, desirable versus undesirable outcomes in games of chance, etc.). 
The relative frequencies and their limits were the obvious entities to work with. But the 
information can be represented equally well by other random variables $\chi_j$, as long as 
these are one-to-one mappings $\chi_j(\nu_j)$, so that no information is lost. The question 
is, whether there exists a most efficient representation.

To answer this, let us see what we know about the limits $p_1,...,p_K$ {\em before} the 
experiment, but having decided to do $N$ trials. Our analysis is equivalent for all $K$ 
outcomes, so that we can pick out one and drop the subscript. We can use Chebyshev's 
inequality\cite{Chebyshev} to estimate the width of the interval, to which the probability 
$p$ of the chosen outcome is pinned down.\footnote{Chebyshev's inequality states: For 
any random variable, whose standard deviation exists, the probability that the value of the 
random variable deviates by more than $k$ standard deviations from its expectation value is 
less than, or equal to, $k^{-2}$. Here, $k$ is a free confidence parameter greater 1.}

 If $N$ is not too small, we get
\be
w_{p} = 2k \sqrt{\frac{\nu(1-\nu)}{N}},
\ee
where $k$ is a free confidence parameter. (Eq.(4) is not valid at $\nu$=0 or 1.) Before the 
experiment we do not know $\nu$, so we can only give the upper limit, 
\be
w_{p} \le \frac{k}{\sqrt{N}}.
\ee
But we can be much more specific about the limit $x$ of the random variable 
$\chi(\nu)$, for which we require that, at least for large $N$, the standard deviation 
$\Delta \chi$ shall be {\em independent} of $p$ (or of $x$ for that matter, since 
there will exist a function $p(x)$),
\be
\Delta \chi = \frac{C}{\sqrt{N}},
\ee
where $C$ is an arbitrary real constant. A straightforward analysis reveals
\be
\chi = C \arcsin\left(2\nu-1\right) + \theta, 
\ee
where $\theta$ is an arbitrary real constant.\cite{Wootters} For comparison with 
$\nu$ we confine $\chi$ to [0,1] and thus set $C =\pi^{-1}$ and $\theta=.5$. 
Then we have $\Delta \chi =1/(\pi \sqrt{N})$, and upon application of 
Chebyshev's inequality we get the interval $w_{x}$ to which we can pin down the 
unknown limit $x$ as
\be
w_{x} =\frac{2}{\pi}\frac{k}{\sqrt{N}}.
\ee
Clearly, this is narrower than the upper limit for $w_{p}$ in eq.(5). Having done no 
experiment at all, we have better knowledge on the value of $x$ than on the value of $p$, 
although both can only be in the interval [0,1]. And note that, the actual experimental data 
will add nothing to the accuracy with which we know $x$, but they may add to the accuracy 
with which we know $p$. Nevertheless, even with data, $w_p$ may still be larger than 
$w_x$, especially when $p$ is around 0.5. Figure 1 shows the relation of $\nu$ and 
$\chi$, and how the probability distribution of $\nu$ for various values of $p$ gets 
squeezed and stretched when plotting it for $\chi$.

For the representation of information the random variable $\chi$ is {\em the} proper 
choice, because it disentangles the two aspects of empirical information: The number of trials 
$N$, which is determined by the experimenter, not by nature, and the actual data, which are 
only determined by nature. The experimenter controls the accuracy $w_x$ by deciding $N$, 
nature supplies the data $\chi$, and thereby the whereabouts of $x$. In the real domain 
the only other random variables with this property are the linear transformations afforded by 
$C$ and $\theta$. From the physical point of view $\chi$ is of interest, because its 
standard deviation is an {\em invariant of the physical conditions} as contained in $p$ or 
$x$. The random variable $\chi$ expresses empirical information with a certain efficiency, 
eliminating a numerical distortion that is due to the structure of the multinomial distribution, 
and which is apparent in all other random variables. We shall call $\chi$ an {\em efficient 
random variable} (ER). More generally, we shall call any random variable an ER, whose 
standard deviation is asymptotically invariant of the limit the random variable tends to, eq.(6).

A graphical depiction of the relation between $\nu$ and $\chi$ can be given by drawing 
a semicircle of diameter 1 along which we plot $\nu$ (Fig.2a). By orthogonal projection 
onto the semicircle we get the random variable $\zeta=[\pi+2\arcsin(2\nu-1)]/4$ 
and thereby $\chi$, when we choose different constants. The drawing also suggests a 
simple way how to obtain a {\it complex} ER. We scale the semicircle by an arbitrary real 
factor $a$, tilt it by an arbitrary angle $\varphi$, and place it into the complex plane as 
shown in Fig.2b. This gives the random variable 
\be
\beta = a \left(\sqrt{\nu\left(1-\nu\right)} + i\nu\right) 
e^{-i\varphi} + b
\ee
where $b$ is an arbitrary complex constant. We get a very familiar special case by setting 
$a=1$ and $b = 0$:
\be
\psi = \left(\sqrt{\nu\left(1-\nu\right)} + i\nu\right) 
e^{-i\varphi}.
\ee
For large $N$ the probability distribution of $\nu$ becomes gaussian, but also that of any 
smooth function of $\nu$, as can be seen in Fig.1. Therefore the standard deviation of 
$\psi$ is obtained as 
\be
\Delta \psi=\left|\frac{\partial \psi}{\partial \nu }\right| 
\Delta \nu = \frac{1}{2\sqrt{N}}.
\ee
Obviously, the random variable $\psi$ is an ER. It fulfills 
$\left|\psi\right|^2=\nu$, and we recognize it as the {\it probability 
amplitude} of quantum theory, which  we would infer from the observed relative frequency 
$\nu$. Note, however, that the intuitive way of getting the quantum mechanical probability 
amplitude, namely, by simply taking $\sqrt{\nu}\exp(i \alpha)$, where $\alpha$ 
is an arbitrary phase, does {\it not} give us an ER.

We have now two ways of representing the obtained information by ERs, either the real 
valued $\chi$ or the complex valued $\beta$. Since the relative frequency of each of the 
$K$ outcomes can be converted to its respective efficient random variable, the result of a 
general probabilistic experiment is efficiently represented by the vector 
($\chi_1$,...,$\chi_K$), or by the vector ($\beta_1$,...,$\beta_K$). The latter is 
equivalent to the quantum mechanical state vector, if we normalize it: 
($\psi_1$,...,$\psi_K$).

At this point it is not clear, whether fundamental science could be built solely on the real ERs 
$\chi_j$ or whether it must rely on the complex ERs $\beta_j$, and for practical reasons 
on the normalized case $\psi_j$, as suggested by current formulations of quantum theory. 
We cannot address this problem here, but mention that working with the $\beta_j$ or 
$\psi_j$ can lead to nonsensical predictions, while working with the $\chi_j$ never 
does, so that the former are more sensitive to inconsistencies in the input data 
\cite{more.in.quantph}. Therefore we use only the $\psi_j$ in the next section, but will 
not read them as if we were doing quantum theory.

\section{Predictions}
Let us now see whether the representation of probabilistic information by ERs suggests 
specific laws for predictions. A prediction is a statement on the expected values of the 
probabilities of the different outcomes of a probabilistic experiment, which has not yet been 
done, or whose data we just do not yet know, on the basis of auxiliary probabilistic 
experiments, which have been done, and whose data we do know. We intend to make a 
prediction for a probabilistic experiment with $Z$ outcomes, and wish to calculate the 
quantities $\phi_s$, ($s=1,...,Z$), which shall be related to the predicted probabilities 
$P_s$ as $P_s = |\phi_s|^2$. We do not presuppose that the $\phi_s$ are ERs. 

We assume we have done $M$ different auxiliary probabilistic experiments of various 
multinomial order $K_m$, $m=1,...,M$, and we think that they provided all the input 
information needed to predict the $\phi_s$, and therefore the $P_s$. With (10) the 
obtained information is represented by the ERs $\psi_j^m$, where $m$ denotes the 
experiment and $j$ labels a possible outcome in it ($j=1,...,K_m$). Then the predictions are
\be
\phi_s = \phi_s(\psi_1^1,...,\psi_{K_1}^1, ... ... ..., 
\psi_1^M,...,\psi_{K_M}^M)
\ee
and their standard deviations are, by the usual convolution of gaussians as approximations of 
the multinomial distributions,
\be
\Delta \phi_s =\sqrt{ \sum_{m=1}^M \frac{1}{4 N_m}\left( 
\sum_{j=1}^{K_m} \left| \frac{\partial \phi_s}{\partial 
\psi_j^m}\right|^2 \right)},
\ee
where $N_m$ is the number of trials of the $m^{th}$ auxiliary experiment. 
If we wish the $\phi_s$ to be ERs, we must demand that the $\Delta \phi_s$ depend 
only on the $N_m$. (A technical requirement is that in each of the $M$ auxiliary 
experiments one of the phases of ERs $\psi_j^m$ cannot be chosen freely, otherwise the 
second summations in (13) could not go to $K_m$, but only to $K_m-1$.)
Then the derivatives in (13) must be constants, implying that the $\phi_s$ are {\em 
linear} in the $\psi_j^m$. 
However, we cannot simply {\em assume} such linearity, because (12) contains the laws of 
physics, which cannot be known {\em a priori}. But we want to point out that a linear 
relation for (12) has very exceptional properties, so that it would be nice, if we found it 
realized in nature. To be specific, if the
$N_m$ are sufficiently large, linearity would afford predictive power, which no other 
functional relation  could achieve: It would be sufficient to know the number of trials of each 
auxiliary probabilistic experiment in order to specify the accuracy of the predicted 
$\phi_s$. No data would be needed, only a decision how many trials each auxiliary 
experiment will be given! Moreover, even the slightest increase of the amount of input 
information, by only doing one more trial in {\em any} of the auxiliary experiments, would 
lead to better accuracy of the predicted $\phi_s$, by bringing a definite decrease of the 
$\Delta \phi_s$. This latter property is absent in virtually all other functional relations 
conceivable for (12). In fact, most nonlinear relations would allow {\em more} input 
information to result in {\em less} accurate predictions. This would undermine the very idea 
of empirical science, namely that, by observation our knowledge about nature can only 
increase, never just stay the same, let alone decrease. For this reason we assume linearity 
and apply it to a concrete example.

We take a particle in a one dimensional box of width $w$.
Alice repeatedly prepares the particle in a state only she knows. At time $t$ after the 
preparation Bob measures the position by subdividing the box into $K$ bins of width $w/K$ 
and checking in which he finds the particle. In $N$ trials Bob obtains the relative frequencies 
$\nu_1,...,\nu_K$, giving a good idea of the particle's position probability distribution at 
time $t$. He represents this information by the ERs $\psi_j$ of (10) and wants to use it to 
predict the position probability distribution at time $T$ ($T>t$).

First he predicts for $t+dt$. With (12) the predicted $\phi_s$ must be linear in the 
$\psi_j$ if they are to be ERs,
\be
\phi_s(t+dt)=\sum_{j=1}^K a_{sj}\psi_j.
\ee
Clearly, when $dt\rightarrow 0$ we must have $a_{sj}=1$ for $s=j$ and 
$a_{sj}=0$ otherwise, so we can write 
\be
a_{sj}(t)=\delta_{sj}+g_{sj}(t)dt,
\ee
where $g_{sj}(t)$ are the complex elements of a matrix ${\bf G}$ and we included the 
possibility that they depend on $t$. Using matrix notation and writing the $\phi_s$ and 
$\psi_j$ as column vectors we have
\be
\vec{\phi}(t+dt) = \left[{\bf 1}+{\bf G}(t)dt\right] \vec{\psi}.
\ee
For a prediction for time $t+2dt$ we must apply another such linear transformation to the 
prediction we had for $t+dt$, 
\be
\vec{\phi}(t+2dt) =\left[{\bf 1}+{\bf G}(t+dt)dt\right] 
\vec{\phi}(t+dt).
\ee
Replacing $t+dt$ by $t$, and using 
$\vec{\phi}(t+dt)=\vec{\phi}(t)+\frac{\partial\vec{\phi}(t)}
{\partial t}dt$,
we have
\be
\frac{\partial\vec{\phi}(t)}{\partial t}= {\bf G}(t) \vec{\phi}(t).
\ee
With (10) the input vector was normalized, $|\vec{\psi}|^2=1$. We also demand this 
from the vector $\vec{\phi}$. This results in the constraint that the diagonal elements 
$g_{ss}$ must be imaginary and the off-diagonal elements must fulfill $g_{sj}=-
g_{js}^*$. And then we have obviously an evolution equation just as we know it from 
quantum theory.

For a quantitative prediction we need to know ${\bf G}(t)$ and the phases $\varphi_j$ 
of the initial $\psi_j$. We had assumed the $\varphi_j$ to be arbitrary. But now we see 
that they influence the prediction, and therefore they attain physical significance. ${\bf 
G}(t)$ is a unitary complex $K \times K$ matrix. For fixed conditions it is independent of 
time, and with the properties found above, it is given by $K^2-1$ real numbers. The initial 
vector $\vec{\psi}$ has $K$ complex components. It is normalized and one phase is 
free, so that it is fixed by $2K-2$ real numbers. Altogether $K^2+2K-3 = (K+3)(K-1)$ 
numbers are needed to enable prediction. Since one probabilistic experiment yields $K-1$ 
numbers, Bob must do $K+3$ probabilistic experiments with different delay times between 
Alice's preparation and his measurement to obtain sufficient input information. But neither 
Planck's constant nor the particle's mass are needed. It should be noted that this analysis 
remains unaltered, if the initial vector $\vec{\psi}$ is obtained from measurement of 
joint probability distributions of several particles. Therefore, (18) also contains entanglement 
between particles.

\section{Discussion}
This paper was based on the insight that under the probabilistic paradigm data from 
observations are subject to the multinomial probability distribution. For the representation of 
the empirical information we searched for random variables which are stripped of numerical 
artefacts. They should therefore have an invariance property. We found as unique random 
variables a real and a complex class of {\em efficient random variables} (ERs). They capture 
the obtained information more efficiently than others, because their standard deviation is an 
asymptotic invariant of the physical conditions. The quantum mechanical probability 
amplitude is the normalized case of the complex class. It is natural that fundamental 
probabilistic science should use such random variables rather than any others as the 
representors of the observed information, and therefore as the carriers of meaning. 

Using the ERs for prediction has given us an evolution prescription which is 
equivalent to the quantum theoretical way of applying a sequence of infinitesimal rotations to 
the state vector in Hilbert space\cite{Lande.Peres.Fivel}. It seems that simply analysing 
how we gain empirical information, what we can say from it about expected future 
information, and not succumbing to the lure of the question what is behind this information, 
can give us a basis for doing physics. This confirms the operational approach to science.
And it is in support of Wheeler's {\em It-from-Bit} hypothesis\cite{Wheeler}, 
Weizs\"acker's {\em ur-theory}\cite{Weizsacker}, Eddington's idea that information 
increase itself defines the rest\cite{Eddington}, Anandan's conjecture of {\em absence 
of dynamical laws}\cite{Anandan}, Bohr and Ulfbeck's hypothesis of {\em mere 
symmetry}\cite{Bohr.Ulfbeck} or the recent {\em 1 Bit --- 1 Constituent} hypothesis of 
Brukner and Zeilinger\cite{Caslav}. 

In view of the analysis presented here the quantum theoretical probability calculus is an 
almost trivial consequence of probability theory, but not as applied to 'objects' or anything 
'physical', but as applied to the naked data of probabilistic experiments. If we continue this 
idea we encounter a deeper problem, namely whether the space which we consider physical, 
this 3- or higher dimensional manifold in which we normally assume the world to unfurl 
\cite{Mohrhoff}, cannot also be understood as a peculiar way of representing data. Kant 
conjectured this - in somewhat different words - over 200 years ago \cite{Kant}. And 
indeed it is clearly so, if we imagine the human observer as a robot who must find a compact 
memory representation of the gigantic data stream it receives through its senses 
\cite{Jaynes}. That is why our earlier example of the particle in a box should only be seen 
as illustration by means of familiar terms. It should not imply that we accept the naive 
conception of space or things, like particles, 'in' it, although this view works well in everyday 
life and in the laboratory --- as long as we are not doing quantum experiments. We think that a 
full acceptance of the probabilistic paradigm as the basis of empirical science will eventually 
require an attack on the notions of spatial distance and spatial dimension from the point of 
view of optimal representation of probabilistic information.

Finally, we want to remark on a difference of our analysis to quantum theory. We have 
emphasized that the standard deviations of the ERs $\chi$ and $\psi$ become 
independent of the limits of these ERs only when we have infinitely many trials. But there is a 
departure for finitely many trials, especially for values of $p$ close to 0 and close to 1. With 
some imagination this can be noticed in Fig.1 in the top and bottom probability distributions of 
$\chi$, which are a little bit wider than those in the middle. But as we always have only 
finitely many trials, there should exist random variables which fulfill our requirement for an ER 
even better than $\chi$ and $\psi$. This implies that predictions based on these 
unknown random variables should also be more precise! Whether we should see this as a 
fluke of statistics, or as a need to amend quantum theory is a debatable question. But it 
should be testable. We need to have a number of different probabilistic experiments, all of 
which are done with only very few trials. From this we want to predict the outcomes of 
another probabilistic experiment, which is then also done with only few trials. Presumably, 
the optimal procedure of prediction will {\em not} be the one we have presented here (and 
therefore not quantum theory). The difficulty with such tests is however that, in the usual 
interpretation of data, statistical theory and quantum theory are treated as separate, while 
one message of this paper may also be that under the probabilistic paradigm the bottom level 
of physical theory should be equivalent to optimal representation of probabilistic information, 
and this theory should not be in need of additional purely statistical theories to connect it to 
actual data. We are discussing this problem in a future paper\cite{long.paper}.

\section*{Acknowledgments}
This paper is a result of pondering what I am doing in the lab, how it can be that in the 
evening I know more than I knew in the morning,  and discussing this with G. Krenn, K. Svozil, 
C. Brukner, M. Zukovski and a number of other people.

\begin{figure}[ht]
\begin{center}
\epsffile{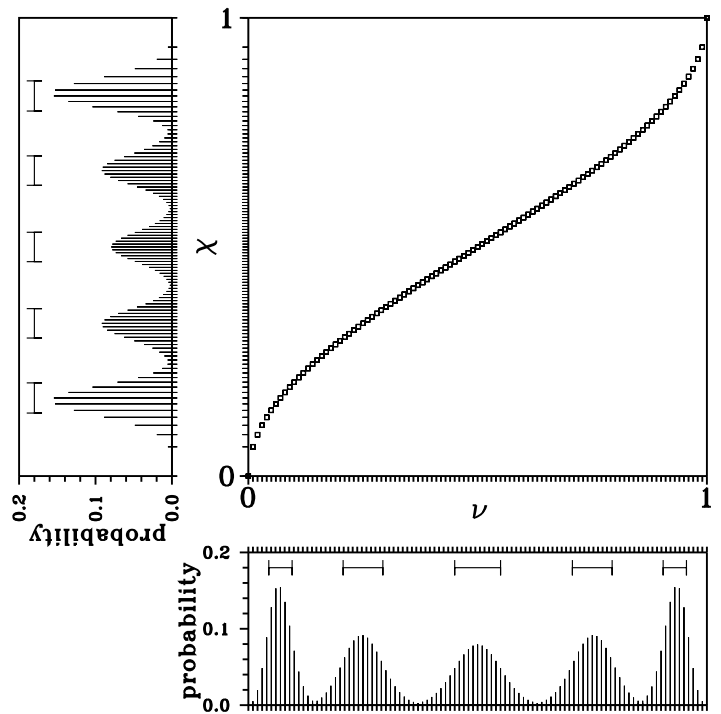}
\caption[Fig.1:]{Functional relation between random variables $\nu$ and $\chi$, 
and their respective probability distributions as expected for $N=100$ trials, plotted for five 
different values of $p$: .07, .25, .50, .75 and .93. The bar above each probablity distribution
indicates twice its standard deviation. Notice that the standard deviations of $\nu$ differ 
considerably for different $p$, while those of $\chi$ are all the same, as required in 
eq.(6)}
\end{center}
\end{figure}

\begin{figure}[ht]
\begin{center}
\epsffile{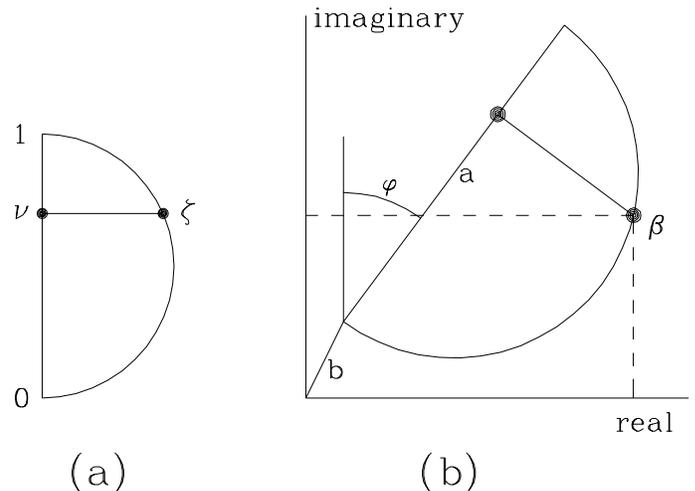}
\caption[Fig.2:]{{\bf (a)} Graphical construction of efficient random variable 
$\zeta$ (and thereby of $\chi$) from the observed relative frequency $\nu$. 
$\zeta$ is measured along the arc. {\bf (b)} Similar construction of the efficient random 
variable $\beta$. It is given by its coordinates in the complex plane. The quantum 
mechanical probability amplitude $\psi$ is the normalized case of $\beta$, obtained by 
setting $a=1$ and $b=0$.}
\end{center}
\end{figure}

\end{document}